\documentclass{ws-procs10x7}
\usepackage{balance}
\usepackage{epsfig}

\newcolumntype{d}[1]{D{.}{.}{#1}}

\newcommand{\tev}{\, {\rm TeV}}
\newcommand{\gev}{\, {\rm GeV}}

\makeindex

\begin{document}

\title{MIXING AND CP-VIOLATION\\
IN THE LITTLEST HIGGS MODEL WITH T-PARITY}
\author{C. TARANTINO$^*$}

\address{Physik Department, Technische Universit\"at M\"unchen,
D-85748 Garching, Germany\\$^*$E-mail: cecilia.tarantino@ph.tum.de}
%%%%%%%%%%%%%%%%%%%%%%%%%%%%%%%%%%%%%%%%%%%%%%%%%%%%%%%%%%%%%%%%%%%%%%%%%
% You may repeat \author \address as often as necessary                 %
%%%%%%%%%%%%%%%%%%%%%%%%%%%%%%%%%%%%%%%%%%%%%%%%%%%%%%%%%%%%%%%%%%%%%%%%%
\twocolumn[\maketitle\abstract{
The results of an extensive flavour physics analysis in the Littlest
Higgs model with T-parity are presented. In particular, mixing and
CP-violation in $K^0-\bar K^0$ and $B^0_{d,s}-\bar B^0_{d,s}$ systems have 
been studied. Several scenarios defined by the values of mirror
quark masses and the parameters of the new flavour mixing matrix $V_{Hd}$ have
been considered. A very interesting scenario is identified, where the 
well-known kaon constraints are satisfied, the $\sin 2 \beta$ and
$\Delta M_s$ issues are improved w.r.t. the Standard Model and visible effects
could be soon found in $A_{SL}^s$ and $S_{\psi \phi}$.
}


]

\section{The Littlest Higgs Model}\label{sec:lhm}
The Standard Model (SM) is in excellent agreement with the
results of particle physics experiments, in particular with
the electroweak (ew) precision measurements, thus suggesting that the SM
cutoff scale is at least as large as $10\tev$.
Having such a relatively high cutoff, however, the SM requires
unsatisfactory fine-tuning to yield a correct ($\approx 10^2\gev$) scale for the Higgs mass, 
being the  Higgs 
boson a fundamental scalar with one-loop quadratically 
divergent corrections to its squared mass.
This ``little hierarchy problem'' has been one of the main motivations to
elaborate models of physics beyond the SM.
While Supersymmetry is at present the leading candidate, 
different proposals have been formulated more recently.
Among them, Little Higgs models play an important role, being
perturbatively computable up to about $10 \tev$ and with a rather small number 
of parameters.

In Little Higgs models\cite{ACG} the Higgs is interpreted as a Nambu-Goldstone 
boson (NGB) corresponding to a spontaneously broken global
symmetry, thus explaining its lightness.
An exact NGB, however, would have only derivative interactions. 
Gauge and Yukawa interactions of the Higgs have to be incorporated. This can
be done without generating
quadratically divergent one-loop contributions to the Higgs mass, through the
so-called {\it collective symmetry breaking}. 
The collective symmetry breaking (SB) has the peculiarity of generating
the Higgs mass only when two or more couplings in the Lagrangian are
non-vanishing.
This mechanism is diagramatically realized through the cancellation of the SM
quadratic divergences by the contributions of new particles with masses around
$1 \tev$.

The most economical, in matter content, Little Higgs model is the Littlest
Higgs (LH)\cite{ACKN}, where the global group $SU(5)$ is spontaneously broken
into $SO(5)$ at the scale $f \approx \mathcal{O}(1 \tev)$ and
the ew sector of the SM is embedded in an $SU(5)/SO(5)$ non-linear
sigma model. 
Gauge and Yukawa Higgs interactions are introduced by gauging the subgroup of
$SU(5)$: $[SU(2) \times U(1)]_1 \times [SU(2) \times U(1)]_2$, with gauge 
couplings respectively equal to $g_1, g_1^\prime, g_2, g_2^\prime$. 
The key feature for the realization of collective SB is that
the two gauge factors commute with a different $SU(3)$ global symmetry
subgroup of $SU(5)$, implying that neither of the gauge factors alone
can generate a potential for the Higgs. 
Consequently, quadratic corrections to the squared Higgs mass involve two
couplings and cannot appear at one-loop.
In the LH model, the new particles appearing at the $\tev$ scales are the heavy
gauge bosons ($W^\pm_H, Z_H, A_H$) the heavy top ($T$) and the scalar triplet 
$\Phi$.

In the LH model, significant corrections to ew observables come
from tree-level heavy gauge boson contributions and the triplet vacuum 
expectation value (vev) which breaks the custodial $SU(2)$ symmetry. 
Consequently, ew precision tests are satisfied for
values of the NP scale $f \ge 2-3 \tev$\cite{HLMW,CHKMT}, too large to solve
the little hierarchy problem.
Motivated by reconciling the LH model with ew precision tests, Cheng and 
Low\cite{CL} proposed to enlarge the symmetry structure of the theory by
introducing a discrete symmetry called T-parity.
T-parity acts as an automorphism which exchanges the $[SU(2) \times SU(1)]_1$ 
and $[SU(2) \times SU(1)]_2$ gauge factors. The invariance of the theory under
this automorphism implies $g_1=g_2$ and $g_1^\prime = g_2^\prime$.
Furthermore, in studied observables, T-parity explicitly forbids the tree-level contributions of  heavy gauge bosons and the
interactions that induced the triplet vev.
The custodial $SU(2)$ symmetry is restored and the compatibility with ew
precision data is obtained already for smaller values of the NP scale, $f \ge
500 \gev$\cite{HMNP}.
Another important consequence is that particle fields are T-even or T-odd
under T-parity. T-even states are the SM particles and the heavy top
$T_+$. T-odd states, instead, are the heavy gauge bosons $W_H^\pm,Z_H,A_H$,
the scalar triplet $\Phi$ and additional particles required by T-parity: 
the odd heavy top $T_-$ and the so-called mirror fermions, i.e.,
fermions corresponding to the SM ones but with opposite T-parity and $\mathcal{O}(1 \tev)$ mass.
Mirror fermions are characterized by new flavour interactions with SM fermions
and heavy gauge bosons, which involve in the quark sector two new unitary 
mixing
matrices analogous to the Cabibbo-Kobayashi-Maskawa (CKM) matrix $V_{CKM}$.
They are $V_{Hd}$ and
$V_{Hu}$, respectively involved when the SM quark is of down- or up-type,
and satisfying $V_{Hu}^\dagger V_{Hd}=V_{CKM}$\cite{L,HLP}. \footnote{$V_{Hd}$
  contains $3$ angles, like $V_\text{CKM}$, but $3$ phases \cite{SHORT}, i.e. 
  two additional phases relative to the CKM matrix.}
The Littlest Higgs model with T-parity (LHT) does not belong to the Minimal
Flavour Violation (MFV) class of models, where the CKM matrix is the only
source of flavour and CP-violation.
The LHT peculiarities are the rather small number of new particles and
parameters (the SB scale $f$, the parameter $x_L$ describing $T_+$ mass and
interactions, the mirror fermion masses and $V_{Hd}$ parameters), the
absence of new operators in addition to the SM ones and, at the same time, the
possibility to have visible effects in flavour observables.

\section{LHT Flavour Analysis} 
Several studies of flavour physics in the LH model without T-parity have been
performed in the last three years\cite{FlavLH}. Without T-parity, mirror 
fermions and new sources of flavour and CP-violation are absent, the LH model
is a MFV model and NP contributions result to be very small. 

More recently, the first flavour physics analysis in the LHT model has been
performed\cite{HLP}, where the mass differences $\Delta M_K$, $\Delta
M_{d,s}$, $\Delta M_D$ and the CP-violation parameter $\varepsilon_K$ have 
been studied.
We have confirmed\cite{BBPTUW} their analytic expressions and performed a
wider phenomenological analysis in the LHT model, including also the width differences 
$\Delta \Gamma_{d,s}$, the radiative decays $B \to X_{s,d} \gamma$ and
the CP-violating asymmetries $A_{CP}(B_d \to \psi K_S)$, $A_{CP}(B_s
\to \psi \phi)$ and $A_{SL}^{d,s}$.
Two interesting issues considered in our analysis are the possible discrepancy
between the values of $\sin 2 \beta$ following directly from $A_{CP}(B_d
\to \psi K_S)$ and indirectly from the usual analysis of the unitarity
triangle\cite{UTfit,CKMfit,BBGT}, and the recent measurement of $\Delta M_s$
by the CDF and D0 collaborations\cite{CDFD0}, that although close to the SM
value is slightly smaller than expected.
An important result of our study is that in the LHT model $\Delta M_s$ can be
smaller than in the SM, as experimentally indicated, and that, for the same
range of parameters, the ``$\sin 2 \beta$'' discrepancy can be cured and
visible effects in observables unknown or with still large uncertainties like
$A_{CP}(B_s \to \psi \phi)$ and $A_{SL}^{d,s}$ can be found.

In our LHT flavour analysis we have considered several scenarios for the
structure of the $V_{Hd}$ matrix and the mass spectrum of mirror
fermions in order to gain a global view over
possible signatures of mirror fermions and T-even contributions.
In all these scenarios the two additional phases of $V_{Hd}$, whose impact is
numerically small, have been set to zero.
The CKM parameters entering the analysis have been taken from tree
level decays only, where NP effects can be neglected.
In order to simplify the numerical analysis we have set all non-perturbative 
parameters to their central values, while allowing $\Delta M_K$, 
$\varepsilon_K$, $\Delta M_d$, $\Delta M_s$, $\Delta M_s / \Delta M_d$ and $S_{\psi K_S}$ to differ from 
their experimental values by $\pm 50\%$, $\pm 40\%$, $\pm 40\%$, $\pm 40\%$,
$\pm 20\%$ and $\pm 8\%$, respectively. This rather conservative choice
guarantees that interesting effects are not missed.
In scenarios $3-5$, described below, the parameters $f$ and $x_L$ have been
fixed to $f=1
\tev$ and $x_L=0.5$ in accordance with ew precision tests.
The main features of the five studied scenarios are described below.

{\bf Scenario 1:}
Mirror fermions are degenerate in mass, therefore
only the T-even sector contributes. This is the MFV limit of the LHT model
where a new phase curing the $\sin 2 \beta$ discrepancy cannot appear and
$(\Delta M_s)_{LHT} \ge (\Delta M_s)_{SM}$, disfavored by the CDF
measurement \cite{CDFD0}, is found.

{\bf Scenario 2:} 
A scan over non-degenerate mirror fermion masses is performed, while $V_{Hd}$ 
has the same structure of the CKM matrix ($V_{Hd}=V_{CKM}$). Similarly to the
previous scenario, no improvements concerning the $\sin 2 \beta$ and $\Delta
M_s$ issues are achieved.

{\bf Scenario 3:}
Mirror fermion masses are
$m_{H1}=400\gev, m_{H2}=500\gev, m_{H3}=600\gev$, while
for $V_{Hd}$ an arbitrary structure is allowed.
The freedom in $V_{Hd}$ allows to soften the $\sin 2 \beta$ discrepancy, while
for $\Delta M_s$ no improvement is found.

{\bf Scenario 4:}
This is the most interesting scenario where the $\Delta M_s$ and $\sin 2 \beta$
issues can be improved and, simultaneously, visible departures from the 
SM and MFV can be obtained, mainly in $A_{SL}^s$ and $A_{CP}(B_s \to
\psi \phi)$. Here
$m_{H1}\approx m_{H2}\approx 500\gev\,, m_{H3}= 1000 \gev$, 
$1/\sqrt{2} \le s_{12}^d \le 0.99$, $ 5\cdot 10^{-5}\le   s_{23}^d \le  2
\cdot 10^{-4}$, $4\cdot 10^{-2}\le  s_{13}^d \le 0.6$
and the phase $\delta^d_{13}$ is arbitrary. The latter sines and phase
are those describing $V_{Hd}$ in a parameterization
  similar to the usual $V_{\rm CKM}$'s one. 
The very different hierarchy of $V_{Hd}$
w.r.t. $V_{\rm CKM}$,
with a large complex phase in the $(V_{Hd})_{32}$ element assures large 
CP-violating effects in the $B^0_s-\bar B^0_s$ system. 
Moreover, $\Delta M_s$ can be smaller than its SM
value and interesting effects in the $B^0_d-\bar B^0_d$ system are found.
The effects in the kaon system are unimportant, being the first two mirror
fermion masses almost degenerate, as required by the
$\Delta M_K$ and $\varepsilon_K$ constraints.

{\bf Scenario 5:}
In all the previous scenarios the SM solution ($71^\circ \pm
16^\circ$) has been considered for the angle $\gamma$ such 
 that only small departures from the SM in the $B^0_d-\bar B^0_d$ system 
 can be consistent with the data. In this last scenario, instead, we have
 assumed the second solution $\gamma = -109^\circ \pm 16^\circ$. 
Moreover,  $m_{H1}=500\gev, m_{H2}= 450\gev, m_{H3}= 1000 \gev$,
$5\cdot 10^{-5}\le s_{12}^d \le 0.015$, $2\cdot 10^{-2}\le s_{23}^d \le 4
\cdot 10^{-2}$, $0.2 \le  s_{13}^d \le 0.5$
and the phase $\delta^d_{13}$ is arbitrary. 
We have found that with this $V_{Hd}$ hierarchy, inverted relative to the CKM
one and also different from that in scenario 4, a rough consistency 
 with the existing data can be obtained.
In spite of that, the combined measurements of $A_\text{SL}^d$ and $\cos 2
\beta$ and the indirect experimental estimate of $A_\text{SL}^s$ make this
scenario very unlikely.
\begin{figure}[b]
\centerline{\epsfig{file=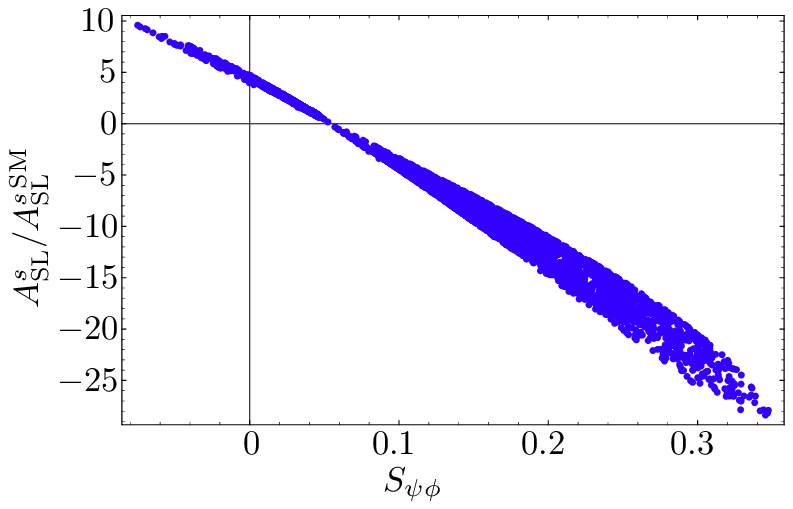,width=2.8in}}
\caption{$A_{\rm SL}^s$ as a function of
  $S_{\psi\phi}$ ($A_{CP}(B_s \to \psi \phi)= S_{\psi\phi} \sin (\Delta M_s t)$), in Scenario 4.}
\label{fig:ASL}
\end{figure}
\begin{figure}[b]
\centerline{\epsfig{file=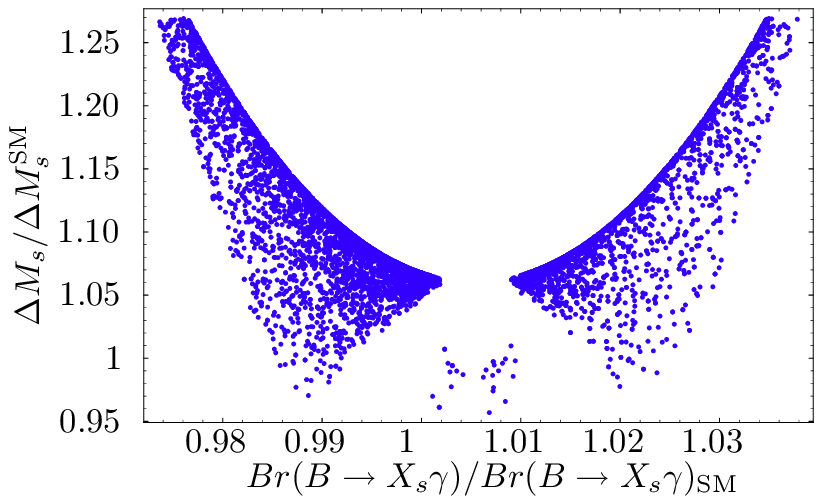,width=2.8in}}
\caption{Correlation between $\Delta M_s$ and $Br(B\to X_s\gamma)$, in Scenario 4.}
\label{fig:Bsg}
\end{figure}

In conclusion, we stress that in the interesting scenario 4, significant 
enhancements of the CP-asymmetries $A_{CP}(B_s \to
\psi \phi)$ ($\approx 5-10$ times larger than in the SM) and 
$A_{SL}^s$ ($\approx 10-20$ times larger than in the SM) are possible as shown 
in Fig.\ref{fig:ASL}, while satisfying all existing constraints, in particular
from the $B \to X_s \gamma$ decay, as shown in Fig.\ref{fig:Bsg}.
In order to improve the LHT flavour analysis the next step is certainly the
inclusion of rare decays in both $K$ and $B_{d,s}$ meson systems.
This study has been just performed\cite{BBPRTUW} finding that visible effects
in observables accessible to future experiments, mainly in $K$ physics,  
are possible.

\section*{Acknowledgments}
I would like to thank the organizers of the interesting and
pleasant conference ICHEP'06, Moscow.
Special thanks go to the other authors of the work presented here:
Monika Blanke, Andrzej J. Buras, Anton Poschenrieder, Selma Uhlig and Andreas 
Weiler. 

\balance


\begin{thebibliography}{20}
\bibitem{ACG}
 N.~Arkani-Hamed {\it et al.},
  Phys.\ Lett.\ B {\bf 513} (2001) 232
  [arXiv:hep-ph/0105239].

\bibitem{ACKN}
  N.~Arkani-Hamed {\it et al.},
  JHEP {\bf 0207} (2002) 034
  [arXiv:hep-ph/0206021].

\bibitem{HLMW}
T.~Han {\it et al.},
Phys.\ Rev.\ D {\bf 67} (2003) 095004
[arXiv:hep-ph/0301040].

\bibitem{CHKMT}
C.~Csaki {\it et al.},
Phys.\ Rev.\ D {\bf 67} (2003) 115002
[arXiv:hep-ph/0211124].
 
\bibitem{CL}
  H.~C.~Cheng and I.~Low,
  JHEP {\bf 0309} (2003) 051 
  [arXiv:hep-ph/0308199];
  JHEP {\bf 0408} (2004) 061
  [arXiv:hep-ph/0405243].

\bibitem{HMNP}
  J.~Hubisz {\it et al.},
  JHEP {\bf 0601} (2006) 135
  [arXiv:hep-ph/0506042].

\bibitem{L}
 I.~Low,
  JHEP {\bf 0410} (2004) 067
  [arXiv:hep-ph/0409025].

\bibitem{HLP}
  J.~Hubisz {\it et al.},
  JHEP {\bf 0606} (2006) 041
  [arXiv:hep-ph/0512169].

\bibitem{SHORT}
  M.~Blanke {\it et al.},
  % ``Another Look at the Flavour Structure of the Littlest Higgs Model with
  %T-Parity,''
  arXiv:hep-ph/0609284.

\bibitem{FlavLH}
  A.~J.~Buras {\it et al.},
  Nucl.\ Phys.\ B {\bf 716} (2005) 173
  [arXiv:hep-ph/0410309];  arXiv:hep-ph/0501230.
  A.~J.~Buras {\it et al.},
  arXiv:hep-ph/0607189.
  S.~R.~Choudhury {\it et al.},
  Phys.\ Lett.\ B {\bf 601} (2004) 164
  [arXiv:hep-ph/0407050].
  J.~Y.~Lee,
  JHEP {\bf 0412} (2004) 065
  [arXiv:hep-ph/0408362].
  S.~Fajfer and S.~Prelovsek,
  Phys.\ Rev.\ D {\bf 73} (2006) 054026
  [arXiv:hep-ph/0511048].
  W.~j.~Huo and S.~h.~Zhu,
  Phys.\ Rev.\ D {\bf 68} (2003) 097301
  [arXiv:hep-ph/0306029].
  S.~R.~Choudhury {\it et al.},
  arXiv:hep-ph/0408125.

\bibitem{BBPTUW}
 M.~Blanke {\it et al.},  arXiv:hep-ph/0605214.
 
\bibitem{UTfit}
  M.~Bona {\it et al.}  [UTfit Collaboration],
  arXiv:\\hep-ph/0509219; arXiv:hep-ph/0605213.
 http://utfit.roma1.infn.it.

\bibitem{CKMfit}
 J.~Charles {\it et al.}  [CKMfitter Group],\\
  Eur.\ Phys.\ J.\ C {\bf 41} (2005) 1
  [arXiv:hep-ph/0406184],\\
  http://www.slac.stanford.edu/xorg/ckmfitter/.

\bibitem{BBGT}
  M.~Blanke {\it et al.},
  arXiv:hep-ph/0604057.

\bibitem{CDFD0}
  A.~Abulencia {\it et al.}  [CDF Collaboration],
  %``Observation of Bs-Bsbar Oscillations,''
  arXiv:hep-ex/0609040.
  D.~Lucchesi  [CDF and D0 Collaborations],
  %``B/s mixing at the Tevatron,''
FERMILAB-CONF-06-262-E
%\href{http://www.slac.stanford.edu/spires/find/hep/www?r=fermilab-conf-06-262-e}{SPIRES entry}
{\it Presented at Workshop on Theory, Phenomenology and Experiments in Heavy Flavor Physics, Capri, May 2006}.

\bibitem{BBPRTUW}
 M.~Blanke {\it et al.}, arXiv:hep-ph/0610298.

\end{thebibliography}
\end{document}